\documentclass[11pt]{article}

\usepackage[final]{acl}

\usepackage{times}
\usepackage{latexsym}
\usepackage{booktabs}
\usepackage{amsmath}
\usepackage{amssymb}
\usepackage{multirow}
\usepackage{pifont}
\usepackage[T1]{fontenc}

\usepackage[utf8]{inputenc}

\usepackage{microtype}

\usepackage{inconsolata}

\usepackage{graphicx}

\title{Prevented is Not Protected: What Attack Success Rate Hides about RAG}

\author{
  \textbf{Chahana Dahal\textsuperscript{1}} 
   \and
  \textbf{Zuobin Xiong\textsuperscript{1}}
\\
  \textsuperscript{1}University of Nevada, Las Vegas 
\\
  \texttt{\{chahana.dahal,zuobin.xiong\}@unlv.edu} 
}

\begin{document}
\maketitle

\newcommand\blfootnote[1]{%
  \begingroup
  \renewcommand\thefootnote{}\footnote{#1}%
  \addtocounter{footnote}{-1}%
  \endgroup
}

\blfootnote{Under review.}

\begin{abstract}

Retrieval-Augmented Generation (RAG) is vulnerable to knowledge poisoning, where adversaries manipulate retrieved evidence to influence model outputs.
Existing defenses are primarily evaluated using attack success rate (ASR), which obscures defense-induced failures and the mechanisms behind apparent attack prevention.
To address this issue, this work introduces \textbf{Aligned RAG Evaluation}, a matched-query protocol that jointly examines clean and poisoned responses, with and without defense.
By combining matched conditions analysis, LLM-judge endorsement, and retrieval dependence controls, the proposed framework distinguishes prevented from introduced failures, target mentions from LLM endorsement, and single-passage influence from evidence composition.
Experiments on HotpotQA and Natural Questions reveal that existing defenses can introduce unexpected target-positive answers, although they can prevent some attacks, while evaluation criteria can also change defense rankings.
Our findings expose critical blind spots in ASR-based evaluation and motivate a more comprehensive assessment of RAG poisoning defenses.
The code and the evaluation artifact are available at \url{https://github.com/chxhana/aligned-rag-eval}.

\end{abstract}

\section{Introduction}
\label{sec:intro}

Retrieval-augmented generation (RAG) allows a language model to generate output using evidence retrieved from an external corpus~\cite{lewis2020retrieval}, which, although effective, also makes the unreliable external corpus part of the model's input.
A RAG poisoning attacker can add crafted passages to the corpus, and those passages can be retrieved at the inference stage and steer the model toward an attacker-chosen answer~\cite{zhong-etal-2023-poisoning, zou2025poisonedrag}.
As a countermeasure, defense mechanisms should therefore reduce the malicious influence introduced by the attacker without degrading the system's performance on clean/benign questions.
However, most defenses of RAG poisoning evaluate whether a defense can \textbf{prevent} an attack from causing the model to answer desired targets~\cite{zou2025poisonedrag,xiang2026certifiably,zhou2025trustrag,zhang2025benchmarking}, yet this approach is not a comprehensive evaluation.
A defense mechanism can change the system's behavior on many queries, including malicious and clean, \emph{by blocking some poisoned answers, introducing new target-positive answers on other poisoned queries, or making the model abstain}.
Thus, a leaderboard that looks at only prevented attacks of a defense mechanism may not reflect the overall safety of the defended system.

To address this issue, this paper introduces \textbf{Aligned RAG Evaluation}, a low-cost yet effective mechanism that uses matched records to evaluate the impact of defense mechanisms.
Particularly, we save four answers for each query to check the prevented and newly introduced attacks: clean retrieval without a defense, clean retrieval with a defense, poisoned retrieval without a defense, and poisoned retrieval with a defense, with all other variables fixed, including query, target, victim model, and decoding policy.
In addition, we also save the retrieved passages and rerun answers after removing these passages, which allows us to check whether the model just mentioned the target, actually endorsed it, or needed both poisoned and legitimate evidence to produce it.
This evaluation protocol is applied to PoisonedRAG passages~\cite{zou2025poisonedrag} on HotpotQA~\cite{yang2018hotpotqa} and Natural Questions (NQ)~\cite{kwiatkowski2019natural}.
Other than the basic evaluation, we also introduce 
\textbf{split-knowledge}, a probe where the target answers appear only when the injected passage is combined with existing legitimate evidence in the corpus.
The enhanced protocol lets us compare two mechanisms: evidence displacement with PoisonedRAG and evidence composition with split-knowledge.

The main results of this work demonstrate the claim: \textbf{prevented is not protected} in RAG defenses from different perspectives.
(i) preventing attacks does not always lower risk.
Under string labels, a defense can block poisoned target outputs while introducing new wrong answers.
(ii) the choice of evaluation metric can substantially change the conclusions about defense.
On the same saved outputs, a defense that appears ineffective under string matching becomes the strongest under LLL-judged endorsement.
(iii) the retrieval score detectors fail when the retrieved context is mostly poisoned, e.g., we found that the detector intercepts 59 of 60 target-positive outputs when there is one poison in the context, but intercepts none with five poisoned.
Together, these findings argue that RAG defenses should be evaluated as interventions on the whole system.
A comprehensive defense report should show what the defense prevented, what it introduced, how it affected clean queries, and whether target strings were actually endorsed.

\begin{figure*}[t]
    \centering
    \includegraphics[width=\textwidth]{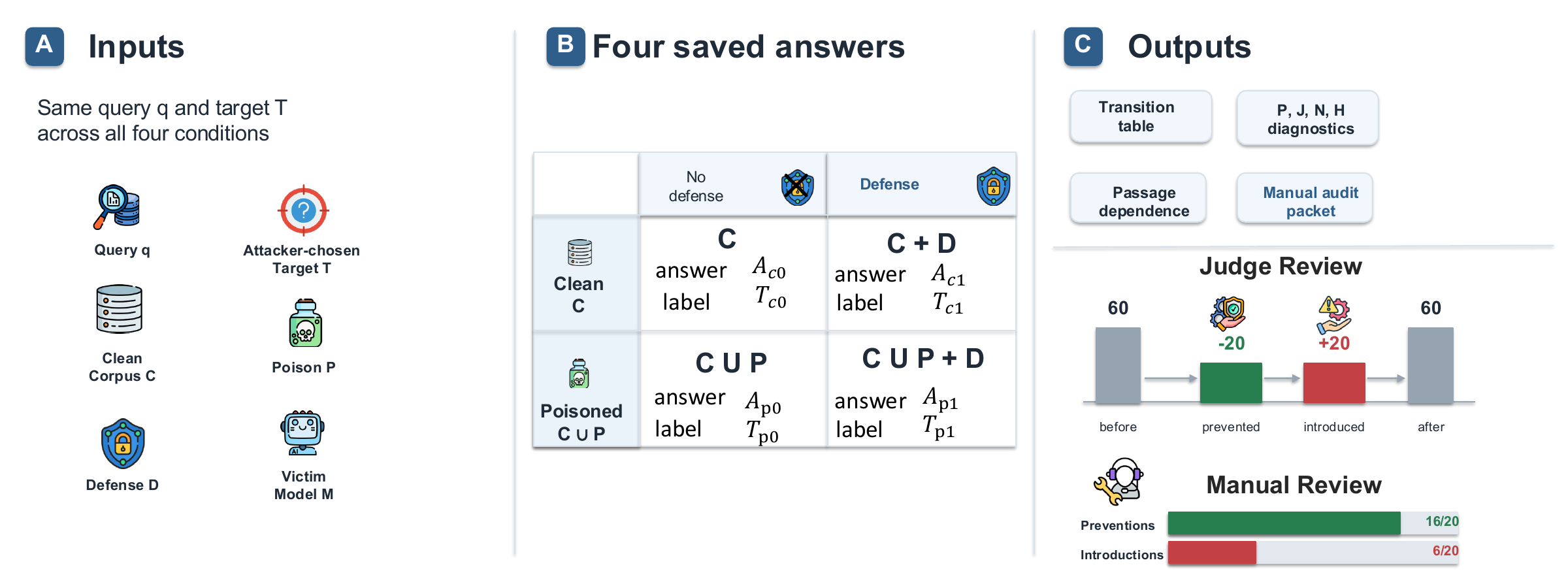}
    \caption{
    Overview of Aligned RAG Evaluation.
    }
    \label{fig:aligned-overview}
\end{figure*}

The contributions of this paper are as follows.
\begin{enumerate}

\item \textbf{An auditable evaluation protocol.}
We define a four-condition record that joins clean and poisoned answers, with and without a defense, and stores retrieval exposure and removal controls for query- and passage-level audit.

\item \textbf{A split-knowledge probe.}
We introduce the split-knowledge probe where the attacker's target depends on combining one injected passage with legitimate corpus evidence, which helps us analyze composition-dependent poisoning from single-passage dominance (\S\ref{sec:res-mechanism}).

\item 
Across HotpotQA and NQ, we show that ASR-style summaries either hide introduced targets (\S\ref{sec:res-prevent}), or depend on the scorer (\S\ref{sec:res-label}), or face detector collapse when poison dominates the retrieved context (\S\ref{sec:res-detector}).

\item
We release \textsc{AlignedRAG-Eval}, a pip-installable harness that applies the protocol to any defense with query-level JSONL outputs, which can produce transition tables, diagnostics, removal-control summaries when available, and audit packets for LLM or human rescoring.

\end{enumerate}


\section{Aligned RAG Evaluation}
\label{sec:method}
\subsection{Four Matched Conditions}
\label{sec:paired}

\footnote{Due to page limit, the Related Works section is moved to Appendix~\ref{sec:rel-work}}
For a RAG poisoning attack, the adversary's target is to make the model answer the malicious or desired output.
For each query $q$, we keep the attacker target $t$ fixed.
We then retrieve two contexts from the target RAG system: a clean context $c$ from the original corpus $\mathcal C$, and a poisoned context $p$ from $\mathcal C \cup \mathcal P$, where $\mathcal P$ is the set of injected passages.
Next, we can generate answers from the retrieved contexts with and without a defense.
The contexts and answers can be organized into a matrix from the four matched conditions, as shown in Table~\ref{tab:four-conditions}, where each entry uses the same query, target, answering model, and decoding policy.

\begin{table}[t]
\centering\small
\begin{tabular}{@{}lll@{}}
\toprule
Retrieval & No defense & Defense \\
\midrule
Clean & $T_{c0}$ & $T_{c1}$ \\
Poisoned & $T_{p0}$ & $T_{p1}$ \\
\bottomrule
\end{tabular}
\caption{Four matched conditions for each query.
$T=1$ means the answer receives a positive target label.}
\label{tab:four-conditions}
\end{table}

We define a \textbf{target-positive} answer according to the evaluation criterion: an answer is target-positive if it contains the attacker's target string under string matching, or if it is semantically endorsed as the target under LLM-based judging.
Let $T_{sd}$=1 indicate answer is target-positive and $T_{sd}$=0 otherwise, where $s\in\{c,p\}$ denotes the retrieval condition: clean ($c$) or poisoned ($p$), and $d\in\{0,1\}$ denotes whether the defense is off ($0$) or on ($1$).
Let $r_{sd}$ be the fraction of queries where $T_{sd}=1$.
For example, $r_{p0}$ is the target-positive rate under poisoned retrieval with no defense, and $r_{p1}$ is the target-positive rate under poisoned retrieval with defense.

We first measure two opposite transitions under poisoned retrieval.
Let Prevention $P$ be the fraction of all queries where the defense changes the answer from target-positive $T_{p0}=1$ to target-negative $T_{p1}=0$.
Let Introduction $J$ be the fraction of all queries where the defense changes the answer from target-negative $T_{p0}=0$ to target-positive $T_{p1}=1$.
The net target reduction is therefore
\begin{equation}
N =P - J=r_{p0} - r_{p1}.
\end{equation}

This difference matters because a defense can prevent some target-positive answers while introducing others in practice.
However, the success attack rate metric reports only the first dimension, with the rate $P/r_{p0}$ when $r_{p0}>0$.
Using the Net reduction $N$ reports the combined effect over the full query set.

We also measure the defense's effect on clean retrieval context for comparison.
The clean-target change is calculated as follows:
\begin{equation}
H = r_{c1} - r_{c0},
\end{equation}
where a positive $H$ means the adopted defense mechanism increases target-positive answers even when retrieval is clean.

Finally, we compute a factorial contrast:
\begin{align}
\Gamma
&= (r_{p0}-r_{c0}) - (r_{p1}-r_{c1}) \nonumber\\
&= N + H.
\end{align}

The factorial contrast measures how much a defense reduces the poisoning-induced increase in target-positive responses relative to clean retrieval. 
However, a positive contrast $\Gamma$ does not necessarily indicate improved security, as it may arise from an undesirable increase in target-positive responses under clean retrieval. 
Therefore, we interpret this metric jointly with net target reduction and clean-target change.
For this reason, we always report $N$, $H$, and $\Gamma$ together with the absolute poisoned-target rate.


\subsection{Retrieval and Passage Dependence}
\label{sec:retrieval-metrics}
The four-condition record shows us how a target answer is changed, and the passage dependence tells us why it changes by testing which retrieved passages the answer depends on. 
To evaluate the dependence, we keep every selected query, including queries where no poison is retrieved.
For each query, we record the retrieved passages, the ratio of poisoned contexts, whether legitimate supporting evidence was present, and the model's answers under passage-removal controls.

A target answer is defined as \textbf{newly induced} if the clean answer is target-negative but the poisoned answer is target-positive.
When removing those poisoned passages, we keep the remaining passages in the same order and do not retrieve any replacements.
Then, an answer is \textbf{injection-dependent} if the target-positive label disappears after we remove the poisoned passages from the context.
Besides, we also run two smaller control experiments:
A poison-only control that keeps only the poison passages and an individual-poison control that keeps one poison passage at a time.
These control experiments show whether the poison works only as a group, or whether one injected passage is enough by itself.

In the split-knowledge probe, the target-positive appears only when the model combines contexts from legitimate evidence in the corpus with injected evidence.
Let $A$ denote the legitimate evidence passages, $B$ the injected passages, and $\mathcal{H}$ the remaining retrieved passages.
We evaluate the split-knowledge probe under four different contexts compositions: $\{A+B+\mathcal{H}$, $A+\mathcal{H}$, $B+\mathcal{H}$, $\mathcal{H}\}$.
In addition, two cases are defined to facilitate the analysis: \textbf{strict case} and \textbf{strict newly induced case}.
A \textbf{strict case} checks whether the target-positive appears only when the model sees both legitimate and injected evidence together.
The case is strict if the answer is target-positive with $\{A+B+\mathcal{H}$ but target-negative with $A+\mathcal{H}$, $B+\mathcal{H}$, $\mathcal{H}\}$.
A \textbf{strict newly induced case} is an enhanced case of the strict case, which requires that the clean answer must be target-negative.
Hence, the strict case is dependent on both $A$ and $B$, while the strict newly induced case introduces a new target-positive answer that was not present during clean retrieval.

\section{Experimental Setup}
\label{sec:exp}

\paragraph{Attacks and Models.}

We evaluate PoisonedRAG~\cite{zou2025poisonedrag} on 100 queries each from HotpotQA~\cite{yang2018hotpotqa} and Natural Questions (NQ)~\cite{kwiatkowski2019natural}, using the corresponding BEIR corpora~\cite{thakur2021beir}.
We additionally evaluate a split-knowledge attack on HotpotQA, where the injected passage induces the target answer only when combined with legitimate retrieved evidence.
For retrieval, we use Contriever~\cite{izacard2021unsupervised} to retrieve the top-$5$ passages in the main experiments.
The split-knowledge experiments additionally evaluate BGE-base English (\texttt{BAAI/bge-base-en-v1.5})~\cite{xiao2024c}.
We use Llama-3-8B-Instruct~\cite{metallamaguard3} as the primary answering model. 
To assess whether our findings generalize across model sizes and families, Llama-3.1-70B-Instruct-AWQ~\cite{lin2024awq} using vLLM and Mistral-7B-Instruct-v0.3~\cite{mistral7b} are also evaluated.

\paragraph{Defenses.}
We evaluate two categories of defenses: detection-based
and pipeline-based defenses.
Detection-based defenses analyze the retrieved context using graph-based features or retrieval-score statistics.
If the resulting detection score exceeds a calibrated threshold, the system abstains from answering.
We evaluate three pipeline-based defenses:
KeywordAgg, adapted from released RobustRAG~\cite{xiang2026certifiably}, 
TrustRAG filtering with conflict resolution~\cite{zhou2025trustrag} and our proposed prompt control baseline, which adds an instruction asking the model to be cautious about unreliable evidence.

We preserve their core defense mechanisms while standardizing context inputs, model calls, decoding, and scoring, which allows all defenses to be evaluated on the same saved retrieval contexts.
We evaluate pipeline-based defenses with the Llama
models, while the Mistral experiments are limited
to detection-based defenses because the pipeline adapters were built for the Llama-family victim.

\paragraph{Metrics and Statistics.}
We evaluate attack outcomes using two scoring methods.
First, we report target-string matching rates for comparability with prior RAG poisoning studies.
Second, we use \texttt{gpt-4o-mini} as an LLM judge ~\cite{zheng2023judging} to reassess the outputs from the main four-condition experiments.
Unlike string matching, the LLM judge distinguishes between merely mentioning the attacker's target and endorsing it as an answer. 
It also evaluates answer correctness against the reference and identifies abstentions.
To assess the LLM judge's reliability, we conduct a blinded human audit and sample additional cases involving label disagreements, defense-induced transitions, or uncertain judgments\footnote{We include the completed audit packet in the supplementary code materials, and the authors provide all audit labels on generated model outputs.}. We also report $kappa$ for human and judge agreement.

We quantify statistical uncertainty using 95\% confidence intervals. 
For paired defense effects, we use nonparametric bootstrap intervals~\cite{tibshirani1993introduction}, clustering resamples by shared injection group.
For individual outcome rates, we report 95\% Wilson score intervals~\cite{wilson1927probable}.
Additional details on retrieval settings, detector calibration, judge prompts, and human audit sampling are provided in \autoref{app:setup-details}.

\begin{figure*}[t]
\centering
\includegraphics[width=0.70\textwidth]{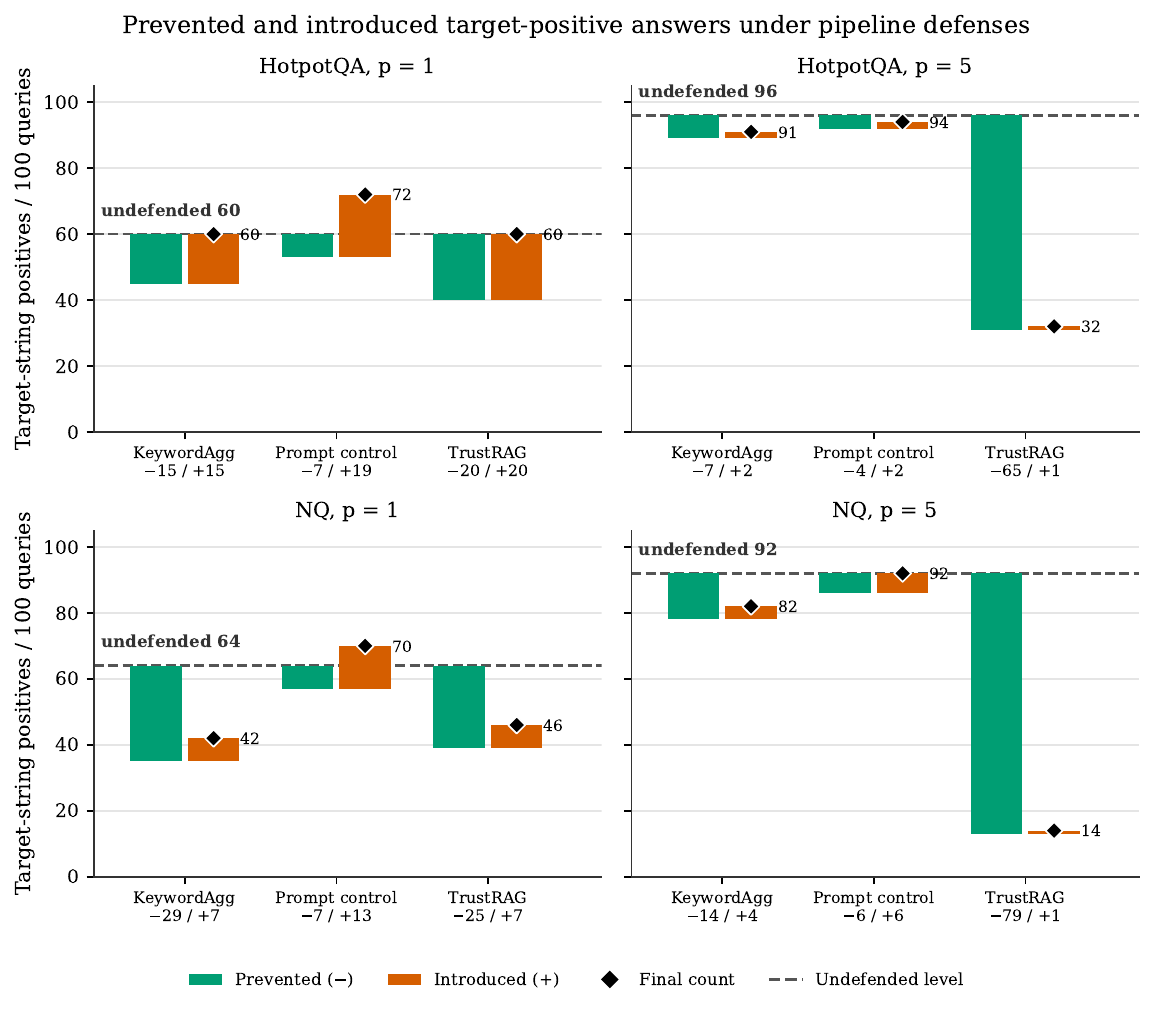}
\caption{%
Pipeline transitions under automatic target-string matching, with 100 queries per panel.
The green bar shows prevented target-positive answers, the orange shows introduced ones, and the diamond marks the total target-positive answers after defense.
Dashed lines show the target-positive without defense. 
}
\label{fig:pipeline-waterfall}
\end{figure*}

\section{Experimental Results}
\label{sec:results}

We structure our evaluation around four research questions (RQs) that examine the limitations of conventional attack success rate (ASR) in evaluating RAG security.
\textbf{RQ1}: Defense Effectiveness.
Does a defense reduce overall attack success and improve system security?
\textbf{RQ2}: Evaluation Reliability.
How do different attack success criteria affect conclusions about defense effectiveness?
\textbf{RQ3}: Detection Robustness.
How does the number of injected passages affect the effectiveness of poisoning detectors?
\textbf{RQ4}: Evidence Dependence.
Does a successful poisoning attack rely on a single injected passage, or on combining poisoned and legitimate evidence?

Together, these questions examine what conventional
ASR-based evaluation overlooks underlying successful attacks.


\subsection{RQ1: Defense Effectiveness}
\label{sec:res-prevent}

We start by evaluating whether a defense reduces target-positive answers under poisoned retrieval.
We compare defended and undefended answers for the same queries using the four-condition record.
\textbf{Prevented} targets are the change from positive to negative, and \textbf{Introduced} targets are the change from negative to positive.
\textbf{Undefended level} is the model performance without the defense, and \textbf{Defended answer} is its answer with defense.
In this analysis, a target-string matching is positive when the answer contains the attacker target as a case-insensitive substring.

Figure~\ref{fig:pipeline-waterfall} shows that prevention does not always mean a lower target-positive rate.
On HotpotQA with one poison passage ($p$=1), the TrustRAG adapter prevents 20 target-positive answers but introduces 20 new, therefore, the final count is unchanged at 60/100.
Similarly, KeywordAgg prevents 15 target-positive answers and introduces 15.
However, at five poison passages ($p$=5), TrustRAG prevents 65 targets and introduces only one, reducing the target-positive from 96 to 32.
On NQ, similar observations are obtained, while the 5 poison samples are largely defended by TrustRAG.
In summary, the prevention and introduction vary across different settings of existing defenses.

Table~\ref{tab:factorial-results} also demonstrates why $\Gamma=N+H$ should not be used alone to compare defenses.
On HotpotQA with one poison, TrustRAG has $\Gamma=12$, but $N=0$ and $H=12$.
The positive contrast comes entirely from more target-positive clean answers, with no reduction in poisoned positives.
On NQ, $\Gamma$ is higher for TrustRAG than KeywordAgg (27 vs.\ 23), although KeywordAgg shows fewer poisoned positives (42 vs.\ 46) and smaller clean-target increase ($H=1$ vs.\ 9).
We therefore interpret $\Gamma$ alongside $N$, $H$, and the absolute poisoned-target rate.

\begin{figure}[t]
\centering
\includegraphics[width=\linewidth]{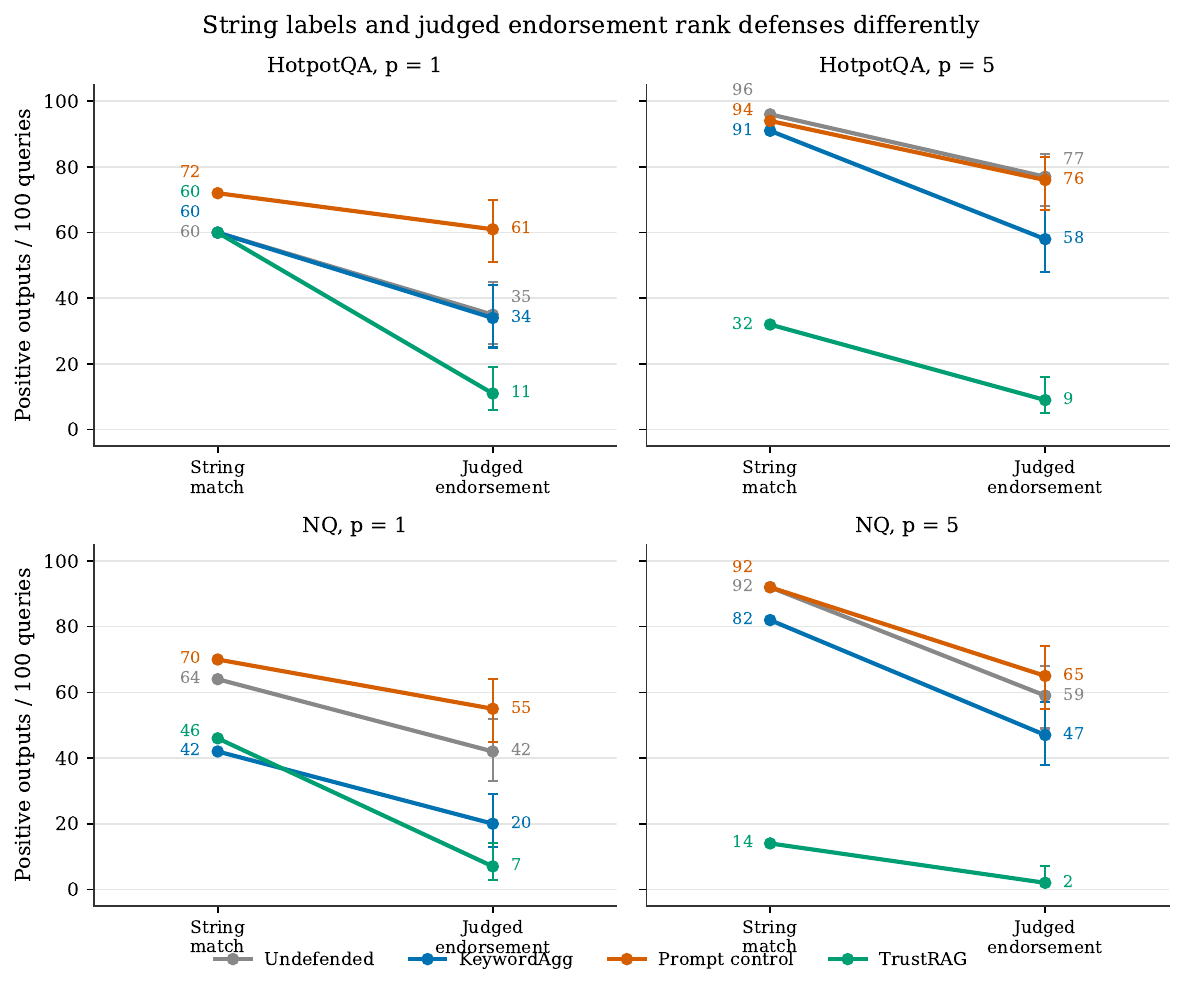}
\caption{%
The same defended and undefended outputs scored in two ways out of 100 queries per panel.
Left points are automatic string-match counts, and right points are independent LLM-judged endorsements.
}
\label{fig:string-vs-judged}
\end{figure}

\begin{table*}[t]
\centering\footnotesize
\begin{tabular}{@{}llrrrr@{}}
\toprule
Dataset & Pipeline & $N$ [95\% CI] & $H$ [95\% CI] & $\Gamma$ [95\% CI] & $p_{\mathrm{McN}}$ \\
\midrule
HotpotQA $p=1$ & KeywordAgg & 0 [$-10$, 9] & $-10$ [$-17$, $-3$] & $-10$ [$-25$, 3] & 1.000 \\
& Prompt control & $-12$ [$-21$, $-4$] & $-13$ [$-20$, $-6$] & $-25$ [$-39$, $-13$] & .029 \\
& TrustRAG & 0 [$-12$, 13] & 12 [5, 21] & 12 [4, 22] & 1.000 \\
\addlinespace
NQ $p=1$ & KeywordAgg & 22 [15, 28] & 1 [0, 3] & 23 [16, 29] & $<.001$ \\
& Prompt control & $-6$ [$-14$, 1] & $-1$ [$-4$, 2] & $-7$ [$-16$, 1] & .263 \\
& TrustRAG & 18 [10, 26] & 9 [6, 12] & 27 [18, 36] & .002 \\
\bottomrule
\end{tabular}
\caption{One-poison diagnostics.
$N$ is poisoned-target reduction, $H$ is clean-target increase, and $\Gamma=N+H$.
}
\label{tab:factorial-results}
\end{table*}

\subsection{Evaluation Reliability}
\label{sec:res-label}

Although TrustRAG prevents 20 target-positive outputs and introduces 20 others under string matching (in HotpotQA, $p$=1), these balanced transitions do not necessarily reflect equivalent changes in actual attack success.
Target-string matching identifies the presence of the attacker's target but cannot distinguish its endorsement or rejection.
To examine this distinction, we manually reviewed all 40 HotpotQA one-poison TrustRAG query pairs whose target-string labels changed.
Among the 20 preventions, 16 were confirmed as genuine preventions, where the undefended answer endorsed the attacker target but the defended answer did not.
In contrast, only 6 of the 20 introductions were confirmed as genuine introductions.
In the remaining 14 cases, the defended answer mentioned the attacker target while questioning or rejecting the poisoned evidence, without endorsing it.
Consequently, the audited transitions reveal 16 genuine preventions and 6 genuine introductions, compared with the balanced 20/20 transitions reported by string matching.
These findings demonstrate that target-string matching can substantially overestimate defense-induced attack success by treating mentions or rejections of poisoned evidence as successful attacks, which undermines the evaluation reliability.

In this section, we score the same saved answers with an independent LLM judge to check whether evaluation methods change the defense ranking.
Figure~\ref{fig:string-vs-judged} shows that target-string matching and LLM judge endorsement give different results for the same answers.
On HotpotQA with one poison, the TrustRAG adapter and the undefended system both have 60 target-string positives.
Under LLM-judged endorsement, TrustRAG has 11 positives compared with 35 without defense.
On NQ with one poison, this target-positive of TrustRAG is higher than KeywordAgg (46 vs. 42) but the LLM-judged endorsement gives a lower score for TrustRAG than KeywordAgg (7 vs. 20).  
The results from Fig.~\ref{fig:string-vs-judged} show that the scoring rule changes defense methods ranking.

\textbf{Remark.} 
The LLM-judged endorsement positive rate includes only answers explicitly labeled as positive by the LLM judge, while uncertain judgments are reported separately.
For example, on HotpotQA with one poisoned passage, the judge assigns uncertain labels to 9 undefended answers and 10 TrustRAG-defended answers as shown in Appendix~\ref{app:judge-label-coverage} (Table~\ref{tab:endorsement-coverage}).
Thus, lower judged endorsement rates should not be interpreted as evidence that all remaining answers are safe.
Unlike TrustRAG, which substantially reduces judged target endorsement, the prompt-only control increases endorsement from 35\% to 61\% on HotpotQA and from 42\% to 55\% on NQ with one poisoned passage.
On HotpotQA, string matching also detects an increase, but its magnitude is smaller (12 vs. 26).
This difference highlights the sensitivity of defense evaluation to the scoring rule.
In particular, string matching treats any mention of the attacker target as a positive outcome, even when the answer explicitly rejects it.
Judged endorsement instead evaluates whether the generated answer semantically supports the attacker's target.

\begin{table}[t]
\centering\footnotesize
\resizebox{\columnwidth}{!}{%
\begin{tabular}{@{}llrrrr@{}}
\toprule
Setting & vs. & $n$ & Rates & Diff. & 95\% CI \\
\midrule
HotpotQA, $p{=}1$ & KeywordAgg & 79 & 11--38 & $-26.6$ & [$-37.5$, $-15.4$] \\
HotpotQA, $p{=}1$ & Prompt     & 80 & 12--66 & $-53.8$ & [$-64.2$, $-42.5$] \\
HotpotQA, $p{=}5$ & KeywordAgg & 79 & 10--71 & $-60.8$ & [$-74.3$, $-47.2$] \\
HotpotQA, $p{=}5$ & Prompt     & 76 & 11--93 & $-82.9$ & [$-93.1$, $-72.0$] \\
NQ, $p{=}1$       & KeywordAgg & 77 & 9--23  & $-14.3$ & [$-24.3$, $-3.9$] \\
NQ, $p{=}1$       & Prompt     & 78 & 9--64  & $-55.1$ & [$-67.1$, $-43.9$] \\
NQ, $p{=}5$       & KeywordAgg & 79 & 3--57  & $-54.4$ & [$-62.7$, $-47.0$] \\
NQ, $p{=}5$       & Prompt     & 65 & 3--91  & $-87.7$ & [$-94.1$, $-80.9$] \\
\bottomrule
\end{tabular}}
\caption{Paired differences in LLM-judged target endorsement on defended poisoned outputs. 
The negative difference (Diff.) means TrustRAG has lower score.}
\label{tab:paired-defense-comparisons}
\end{table}

Table~\ref{tab:paired-defense-comparisons} also compares defenses directly on same set of queries.
For example, on NQ with one poison, 77 queries receive decisive labels (yes/no) for both TrustRAG and KeywordAgg.
Within this subset, TrustRAG has 14.3 percentage points lower judged target endorsement, with a group-bootstrap interval of [$-24.3$, $-3.9$].
The difference is larger at five poisons.
TrustRAG also has lower judged endorsement than the prompt control across both datasets and poison budgets.

\begin{figure}[t]
\centering
\includegraphics[width=\columnwidth]{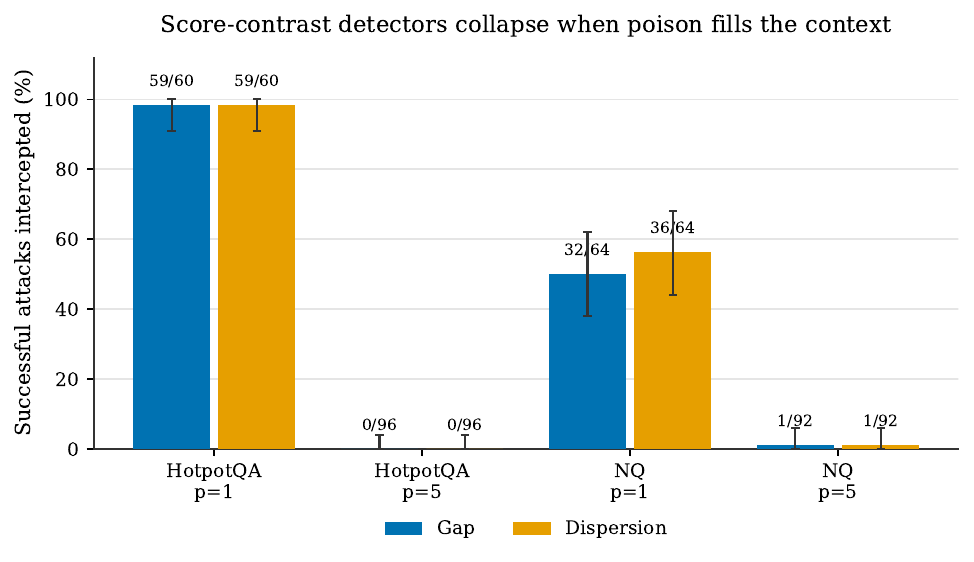}
\caption{%
Target-positive outcomes intercepted at nominal 1\% calibration FPR, with 95\% Wilson intervals for the interception rate.
}
\label{fig:detector-collapse}
\end{figure}

\paragraph{Agreement with Human Judgments.}
We additionally assess the reliability of the LLM judge through a blinded human audit of 100 outputs in Table~\ref{tab:human-judge-agreement}.
Human annotators provide binary yes/no labels, whereas the judge may additionally return \emph{uncertain}.
For target endorsement, five outputs have no defined attacker target and are therefore labeled \emph{not applicable (N/A)} by both evaluators.
Among the remaining 95 outputs, the judge agrees with human annotations on 64, yielding an overall agreement of 69/100 when including the 5 N/A cases and treating uncertain labels as disagreements.

The results show that agreement is substantially higher in decisive yes/no judgments: 64/69 for target endorsement and 67/71 for abstention.
When uncertain judgments are counted as disagreements, the overall agreement is 50/98 for answer correctness and 67/99 for abstention.
These findings support the use of judged endorsement as a complementary evaluation metric, while highlighting the need to report uncertain cases separately.

\paragraph{Reduced Endorsement Does Not Imply Preserved Utility.}
A defense may reduce target endorsement by abstaining or generating vague responses rather than providing correct answers.
To examine this trade-off, we evaluate clean-answer correctness on 200 held-out queries per dataset (Details are reported in Appendix~\ref{app:clean-utility-diagnostics}).
The LLM judge returns uncertain correctness labels for 147 of 200 TrustRAG responses on HotpotQA and 139 of 200 on NQ.
These high uncertainty rates prevent us from drawing reliable conclusions about clean-answer utility.
Thus, although TrustRAG substantially reduces judged target endorsement, whether it preserves answer quality on benign queries remains unresolved.

\subsection{Detection Robustness}
\label{sec:res-detector}
In this section, we investigate how the effectiveness of detection-based defenses changes as the number of injected passages increases.
These defenses analyze the retrieved context before answer generation and trigger abstention when the detection score exceeds a threshold calibrated on clean retrieval contexts.
We evaluate graph-based and retrieval-score-based detectors, with particular emphasis on score-gap and score-dispersion statistics.

Figure~\ref{fig:detector-collapse} reveals a substantial decline in detection effectiveness as the number of injected passages increases on both datasets.
On HotpotQA, both the score-gap and score-dispersion detectors intercept 59 of 60 target-positive outputs with one injected passage ($p=1$), at a 1\% false-positive rate.
However, when the number of injected passages increases to five ($p=5$), both detectors fail to intercept any of the 96 target-positive outputs under the same calibrated thresholds.
A similar pattern emerges on NQ, where the score-gap and score-dispersion detectors intercept 32/64 and 36/64 target-positive outputs at $p=1$, respectively, but only 1/92 each at $p=5$. 
These results demonstrate that detection effectiveness can deteriorate sharply as poisoned passages increasingly dominate the retrieved context.


To examine whether this degradation arises from retrieval encoding artifacts, we apply the same 512-token limit to both legitimate and injected passages.
We additionally compare dot-product scoring with cosine similarity to control for differences in embedding norms.
The detection collapse persists at $p=5$ under these settings, suggesting that unequal token limits and embedding norms do not explain the observed failure.

\paragraph{The Generalization of Detection Failure.}
To examine whether the observed detection failure is specific to the victim model, we repeat the experiments using Mistral-7B-Instruct-v0.3 (Appendix~\ref{app:mistral}, Figure~\ref{fig:mistral-detector-transfer}).
From the results, the attack remains effective, with target-positive rates increasing from 19\% on clean HotpotQA queries to 80\% and 100\% under one and five injected passages, respectively.
On NQ, the corresponding rates increase from 5\% to 72\% and 98\%. 
Despite the high attack success rates, the score-gap detector's interception rate drops sharply as the poison budget increases.
At a 1\% false-positive rate, interception decreases from 78/80 to 0/100 on HotpotQA and from 38/72 to 1/98 on NQ.
These results indicate that the degradation of score-gap-based detection under higher poisoning budgets is not specific to Llama-3-8B-Instruct, but also generalizes to Mistral-7B-Instruct-v0.3.


\subsection{Evidence Dependence}
\label{sec:res-mechanism}

In this section, we investigate whether successful poisoning attacks depend on multiple injected passages or can be triggered by a single passage, since high attack success rates alone cannot distinguish these mechanisms.
To examine this root cause, we conduct removal controls on poisoned target-positive cases: we regenerate responses after removing all injected passages and then with each injected passage retained individually.

\paragraph{Single-Passage Sufficiency.}
Table~\ref{tab:poison-dependence} shows that a single injected passage is sufficient to reproduce the target-positive label in most PoisonedRAG cases.
On NQ with 5 injected passages, 90/100 queries change from target-negative under clean retrieval to target-positive under poisoned retrieval. 
Among these newly induced cases, 87 revert to target-negative when all injected passages are removed.
Moreover, among all 92 poisoned target-positive cases, at least one injected passage independently produces a target-positive response in 91 cases.
This pattern also holds across the additional victim models: a single passage suffices for all 95 target-positive cases on HotpotQA with Llama-3.1-70B (Table~\ref{tab:victim70b}) and for at least 94\% of target-positive cases across the evaluated Mistral-7B settings (Appendix~\ref{app:mistral}).
These findings suggest that most newly induced target-positive outcomes depend on injected evidence, but rarely require multiple injected passages to act jointly.

\begin{table}[t]
\centering\small
\begin{tabular}{@{}lrrrr@{}}
\toprule
& \multicolumn{2}{c}{HotpotQA} & \multicolumn{2}{c}{NQ} \\
\cmidrule(lr){2-3}\cmidrule(l){4-5}
Poison budget $p$ & 1 & 5 & 1 & 5 \\
\midrule
Poisoned positives & 60 & 96 & 64 & 92 \\
Newly induced & 46 & 79 & 62 & 90 \\
Injection-dependent & 45 & 74 & 60 & 87 \\
Single-passage sufficient & 57/60 & 96/96 & 60/64 & 91/92 \\
\bottomrule
\end{tabular}
\caption{PoisonedRAG evidence dependence with top-5 retrieval and Llama-3-8B under target-string labels.
}
\label{tab:poison-dependence}
\end{table}

\begin{table}[t]
\centering
\scriptsize
\setlength{\tabcolsep}{3.5pt}
\begin{tabular}{@{}lrrrr@{}}
\toprule
Victim & $p$ & Clean & Poisoned & Single suff. \\
\midrule
Llama-3-8B & 1 & 17 & 60 & 57/60 \\
Llama-3.1-70B-AWQ & 1 & 8 & 57 & 54/57 \\
Llama-3-8B & 5 & 17 & 96 & 96/96 \\
Llama-3.1-70B-AWQ & 5 & 8 & 95 & 95/95 \\
\bottomrule
\end{tabular}
\caption{HotpotQA PoisonedRAG under victim scaling.
}
\label{tab:victim70b}
\end{table}

\paragraph{Split-Knowledge and Evidence Composition.}
We further investigate whether attacks can induce responses that require combining legitimate evidence $A$ with an injected passage $B$.
We defined the \emph{strict case} as one in which the target appears when both $A$ and $B$ are present, but not when either passage or both are removed, while all background passages remain unchanged.
Table~\ref{tab:split-victim-retriever} shows the results.
Contriever and Llama-3-8B produce target-positive responses for 27 of 311 split-knowledge candidates, of which 23 are newly induced.
However, only 7 satisfy the strict composition criterion, whereas 17 of the 27 target-positive cases can be triggered by an injected passage alone.
We identify two strict cases with BGE cosine retrieval and none with Llama-3.1-70B. 
Although the larger model remains vulnerable to PoisonedRAG, it exhibits no strict split-knowledge successes in our experiments.
Together, these results distinguish attacks driven by individually sufficient poisoned passages from those requiring the combination of legitimate and injected evidence.
Split-knowledge thus serves as a diagnostic for evidence composition rather than evidence of greater attack effectiveness.

\begin{table}[t]
\centering\footnotesize
\setlength{\tabcolsep}{3pt}
\resizebox{\columnwidth}{!}{%
\begin{tabular}{@{}lrrrrrr@{}}
Setting & Clean & Poisoned & New & Strict & Single suff. & Clean EM \\
\midrule
Contriever / Llama-3-8B & 10 & 27 & 23 & 7 & 17/27 & 53 \\
BGE cosine512 / Llama-3-8B & 8 & 18 & 11 & 2 & 12/18 & 66 \\
Contriever / Llama-3.1-70B-AWQ & 8 & 18 & 16 & 0 & 14/18 & 111 \\
\bottomrule
\end{tabular}}
\caption{Split-knowledge attack on 311 fixed HotpotQA candidates.
}
\label{tab:split-victim-retriever}
\end{table}

\section{Conclusion}
In this work, we evaluated the reliability of RAG poisoning defenses other than just looking at the
attack success rates. 
In addition, we introduce Aligned RAG Evaluation, which compares clean and poisoned responses with and without defense on matched queries to reveal both prevented and defense-induced failures.
Experiments on HotpotQA and NQ show that defenses can introduce new target-positive answers, prompt-only controls may increase attacker-target endorsement, and retrieval-score detectors can fail when poisoned passages dominate the retrieved context.
Our evaluation framework provides a more comprehensive
assessment of defense effectiveness, reliability, and
evidence dependence beyond aggregate ASR.

\section*{Limitations}

The PoisonedRAG cells contain 100 queries each, and queries share injected passages in groups of ten.
We account for this by using group-level bootstrap intervals and paired tests, rather than treating every query as fully independent.
The defenses should be read as evaluations of our adapted KeywordAgg, TrustRAG, and prompt-control pipelines, not as exact reproductions of the original papers.
This limitation does not affect the central methodological claim that a defense cannot be judged by prevented attacks alone, because protection also depends on what the defense introduces.

\section*{Ethics Statement}

This work studies retrieval-corpus poisoning to improve RAG defense evaluation.
We use existing benchmarks and released artifacts, report aggregate results, and do not attack live systems.
We release code and records to support reproducible defense auditing.


\bibliography{anthology,custom}

\appendix

\section{Related Works}
\label{sec:rel-work}

\paragraph{Corpus poisoning attacks on RAG.}
Retrieval-augmented generation makes the retrieval corpus part of the model input, so an attacker who can add documents to the corpus can influence downstream answers.
Early corpus-poisoning work showed that a small number of adversarial passages can be retrieved for many queries~\cite{zhong-etal-2023-poisoning}.
PoisonedRAG made this threat targeted.
Each injected passage is crafted for a specific query and asserts an attacker-chosen answer~\cite{zou2025poisonedrag}.
Later work studies related variants, including blocker documents that disrupt retrieval~\cite{shafran2025machine}, single-document attacks~\cite{chang2025one}, and poisoning in graph-structured retrieval~\cite{liang2026graphrag}.
Indirect prompt injection is a neighboring threat where retrieved content contains instructions rather than false evidence~\cite{perez2022ignore,greshake2023not}.
These papers define the attacks and threat models we evaluate.
But we ask what standard success metrics actually measure once a defense is applied.
Our split-knowledge probe also differs from prior poisoning attacks because the injected passage is designed to be insufficient by itself and to work only when combined with legitimate evidence.

\paragraph{Defenses.}
RAG defenses try to identify, isolate, or reduce the influence of untrusted retrieved passages.
RobustRAG answers from isolated passages or passage groups and aggregates the resulting answers and our KeywordAgg adapter follows its keyword-based variant~\cite{xiang2026certifiably}.
TrustRAG filters suspicious retrieved content and then uses model self-assessment to resolve inconsistencies between retrieved and parametric knowledge~\cite{zhou2025trustrag}.
InstructRAG and related prompting methods ask the model to reason explicitly about which passages to trust~\cite{wei2025instructrag}; our prompt-only control is a minimal version of this idea.
Layered systems such as RAGuard combine retriever-side hardening with generation-time counterfactual filtering~\cite{kumar2026raguard}.
These defenses are usually evaluated by counting how many known attacks they stop.
We evaluate the same kind of question, but also ask what the defense introduces, how it changes clean answers, and whether target strings are actually endorsed.

\paragraph{Benchmarks and evaluation protocols.}
Recent benchmarks evaluate RAG poisoning attacks and defenses across datasets, retrievers, and models~\cite{zhang2025benchmarking}.
AgentDojo similarly evaluates security and utility together for prompt-injection defenses in agentic settings~\cite{debenedetti2024agentdojo}.
These benchmarks report attack success and utility as aggregate rates.
Aligned RAG Evaluation is complementary as it saves the four matched answers for each query, so prevention, introduction, clean-side effects, and passage dependence can be inspected and rescored after generation.
This aligns with dynamic evaluation, where the goal is to examine the full results rather than just look at a single score~\cite{kiela2021dynabench}.

\paragraph{Attribution and passage dependence.}
ContextCite attributes generations to context spans and includes poisoning detection as an application~\cite{cohen2024contextcite}.
RAGOrigin assigns responsibility to database texts for misleading generations~\cite{zhang2026taught}, and TRACE detects poisoning through token-influence attribution~\cite{chen2026tracing}.
Our removal, poison-only, and individual-poison controls use the same broad ablation idea, but for a different purpose.
We use simple controls to label outcomes such as whether a target depends on injected passages, whether one injected passage is sufficient, and whether a split-knowledge case truly requires combining poisoned and legitimate evidence.

\section{Mistral-7B Victim Check}
\label{app:mistral}

We repeat the PoisonedRAG attack with Mistral-7B-Instruct-v0.3, keeping the retriever, corpora, injected passages, calibration sets, and nominal 1\% FPR thresholds fixed.
The attack transfers as target-positive rates rise from 19\% clean to 80\% and 100\% on HotpotQA, and from 5\% clean to 72\% and 98\% on NQ, for one and five poisons.
As with Llama-3, nearly all poisoned positives are single-poison sufficient.

\begin{figure}[t]
\centering
\includegraphics[width=.95\linewidth]{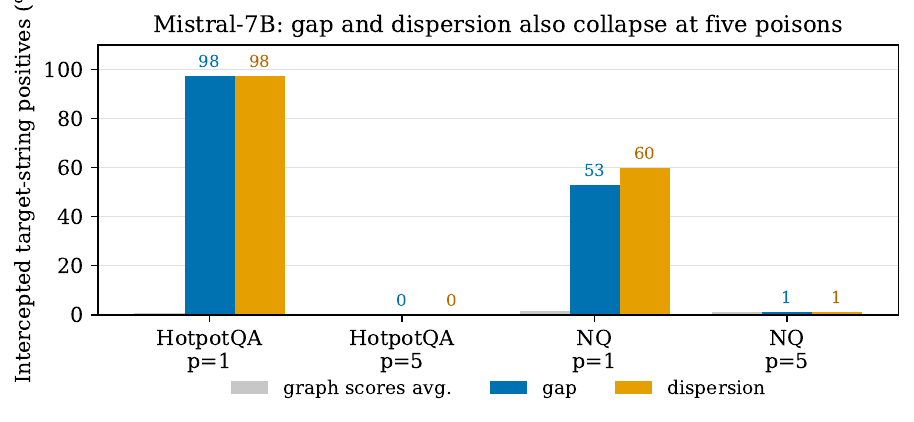}
\caption{
Mistral-7B detector-only check at the nominal 1\% calibration point.
}
\label{fig:mistral-detector-transfer}
\end{figure}

Figure~\ref{fig:mistral-detector-transfer} shows that the detector pattern also transfers.
Gap and dispersion have useful signal at one poison, but collapse at five poisons.
Every detector intercepts at most one target-positive output in the five-poison cells.
We did not rerun pipeline defenses for Mistral, so this check supports the attack-mechanism and detector results, not the pipeline-defense ranking.

\section{Preliminary Clean-Utility Diagnostics}
\label{app:clean-utility-diagnostics}

Table~\ref{tab:judged-clean-utility} reports independent-judge diagnostics on 200 held-out clean queries per dataset.
We keep this result in the appendix because human agreement is weaker for reference correctness than for target endorsement or abstention.
The table is useful for spotting possible utility costs, but we do not use it as a headline defense ranking.

\begin{table*}[t]
\centering\small
\begin{tabular}{@{}llrrrr@{}}
\toprule
Dataset & Method & Ref. yes / resolved & Uncertain & Correct lost & Correct retained \\
\midrule
HotpotQA & No defense & 87/120 & 80 & 0 & 87 \\
HotpotQA & KeywordAgg & 75/96 & 104 & 5 & 59 \\
HotpotQA & Prompt control & 74/104 & 96 & 5 & 65 \\
HotpotQA & TrustRAG & 29/53 & 147 & 5 & 20 \\
\addlinespace
NQ & No defense & 92/134 & 66 & 0 & 92 \\
NQ & KeywordAgg & 71/123 & 77 & 17 & 56 \\
NQ & Prompt control & 83/144 & 56 & 13 & 70 \\
NQ & TrustRAG & 45/61 & 139 & 4 & 23 \\
\bottomrule
\end{tabular}
\caption{Independent-judge clean utility diagnostics on 200 held-out clean queries per dataset.
}
\label{tab:judged-clean-utility}
\end{table*}

\section{Implementation and Annotation Details}
\label{app:setup-details}

\paragraph{Retrieval and budgets.}
PoisonedRAG queries are grouped in tens; a budget $p$ gives each group $10p$ injected passages.
The main setting uses top-5 retrieval with Contriever, corpus passages encoded at 128 tokens, and queries/poisons at 512 tokens with dot-product scores.
We also run two encoding controls: all passages at 512 tokens with dot product, and all passages at 512 tokens with cosine similarity.

\paragraph{Detectors.}
We evaluate six detector scores: spectral connectivity, conductance, modularity, minimum retrieval score, score gap, and score dispersion.
Each detector is calibrated on 500 held-out clean queries to a nominal false-positive rate, then evaluated on 200 clean held-out queries and the attack cells.
Flagged contexts are hard-abstained.
Thresholds are recalibrated per dataset and encoding arm, but shared across poison budgets within the same arm.

\paragraph{Pipelines.}
We adapt released components of RobustRAG and TrustRAG to a common saved-context interface.
RobustRAG uses KeywordAgg and a prompt-only control; TrustRAG uses filtering plus conflict resolution.
The source revisions are pinned: RobustRAG \texttt{9bc35b2}, TrustRAG \texttt{11dcea0}, and PoisonedRAG \texttt{f660d72}.
These are component adaptations under one victim interface, not exact reproductions of every original system detail.

\paragraph{Independent judge and human audit.}
Each distinct saved answer is scored by \texttt{gpt-4o-mini-2024-07-18}.
The judge sees only the question, target, reference aliases with provenance, and answer; it does not see the method or condition.
It labels target endorsement, reference correctness, and abstention, with \emph{uncertain} allowed for each.
The blinded human audit contains 100 outputs, oversampling label flips, TrustRAG cases, prompt-control worsening, and judge-uncertain examples.
Annotators do not see the model, method, condition, or sampling reason.

\paragraph{Uncertainty.}
Paired defense effects use 95\% bootstrap intervals over shared injection groups.
Endpoint rates use 95\% Wilson intervals.

\section{Rates Behind the Four-Condition Contrasts}
\label{app:factorial-rates}
Table~\ref{tab:factorial-absolute} gives the absolute target-positive counts underlying Table~\ref{tab:factorial-results}; Table~\ref{tab:factorial-five} gives the five-poison decomposition. Clean columns repeat across budgets because both budgets share the same clean inputs. The reversal discussed in Finding 1 is visible directly: TrustRAG's defended-clean count on HotpotQA rises from 17 to 29, which is the entire source of its positive one-poison interaction.

\paragraph{Compute.}
All experiments were run on a high-performance computing node with NVIDIA L40S GPUs, each with 48 GB of GDDR6 memory.

\begin{table}[t]
\centering
\footnotesize
\setlength{\tabcolsep}{4.5pt}
\begin{tabular}{@{}llccc@{}}
\toprule
Data & Pipeline & Clean & $p=1$ & $p=5$ \\
\midrule
Hotpot & KeywordAgg & 17 $\to$ 7  & 60 $\to$ 60 & 96 $\to$ 91 \\
       & Prompt     & 17 $\to$ 4  & 60 $\to$ 72 & 96 $\to$ 94 \\
       & TrustRAG   & 17 $\to$ 29 & 60 $\to$ 60 & 96 $\to$ 32 \\
\addlinespace
NQ     & KeywordAgg & 2 $\to$ 3   & 64 $\to$ 42 & 92 $\to$ 82 \\
       & Prompt     & 2 $\to$ 1   & 64 $\to$ 70 & 92 $\to$ 92 \\
       & TrustRAG   & 2 $\to$ 11  & 64 $\to$ 46 & 92 $\to$ 14 \\
\bottomrule
\end{tabular}
\caption{Target-positive counts (out of 100) before $\to$ after defense (automatic string label).}
\label{tab:factorial-absolute}
\end{table}

\begin{table}[t]\centering\small
\begin{tabular}{@{}llrrr@{}}\toprule
Data & Pipeline & $N$ [95\% CI] & $H$ & $\Gamma$ \\ \midrule
Hotpot & KeywordAgg & 5 [0, 11] & $-10$ & $-5$ \\
& Prompt & 2 [$-3$, 7] & $-13$ & $-11$ \\
& TrustRAG & 64 [55, 73] & 12 & 76 \\
\addlinespace
NQ & KeywordAgg & 10 & 1 & 11 \\
& Prompt & 0 & $-1$ & $-1$ \\
& TrustRAG & 78 & 9 & 87 \\
\bottomrule\end{tabular}
\caption{Five-poison factorial decomposition in percentage points ($\Gamma=N+H$) under the automatic target-string label.}
\label{tab:factorial-five}
\end{table}

\section{Judge Label Coverage and Human Agreement}
\label{app:judge-label-coverage}
We report how often the judge returned positive and uncertain target-endorsement labels on the defended poisoned endpoint in Table ~\ref{tab:endorsement-coverage}. Table~\ref{tab:human-judge-agreement} reports agreement with blinded human audit.

\begin{table}[t]
\centering\scriptsize
\setlength{\tabcolsep}{3.5pt}
\begin{tabular}{@{}llrrrr@{}}
\toprule
Data & $p$ & None & Key. & Prompt & Trust \\
\midrule
HotpotQA & 1 & 35/9 & 34/13 & 61/10 & 11/10 \\
HotpotQA & 5 & 77/10 & 58/17 & 76/19 & 9/6 \\
NQ & 1 & 42/11 & 20/13 & 55/13 & 7/11 \\
NQ & 5 & 59/25 & 47/13 & 65/29 & 2/9 \\
\bottomrule
\end{tabular}
\caption{Judge-positive/uncertain target-endorsement counts on poisoned retrieval. Each cell is yes/uncertain out of 100.}
\label{tab:endorsement-coverage}
\end{table}

\begin{table}[t]
\centering\scriptsize
\setlength{\tabcolsep}{2.5pt}
\begin{tabular}{@{}lrrrr@{}}
\toprule
Label & Valid/Unc. & Overall & Decisive \\
\midrule
Target & 95+(5 N/A) / 26 & 69/100, $\kappa=.39$ & 64/69, $\kappa=.74$ \\
Reference & 98 / 30 & 50/98, $\kappa=.20$ & 50/68, $\kappa=.40$ \\
Abstention & 99 / 28 & 67/99, $\kappa=.26$ & 67/71, $\kappa=.72$ \\
\bottomrule
\end{tabular}
\caption{Human-judge agreement in the 100 output blinded audit. Overall counts uncertain labels as disagreement while decisive only counts yes/no. }
\label{tab:human-judge-agreement}
\end{table}

\section{Mechanism Counts}
\label{app:full-counts}

Table~\ref{tab:full-counts} reports the removal-control diagnostics for all completed PoisonedRAG cells.
Across settings, most poisoned positives are reproduced by a single injected passage.
This means the attack usually succeeds through single-passage dominance rather than composition with legitimate evidence.

\begin{table*}[t]\centering\small
\begin{tabular}{@{}lrrrrrr@{}}\toprule
& \multicolumn{4}{c}{HotpotQA $(p,k)$} & \multicolumn{2}{c}{NQ $(p,k)$} \\
\cmidrule(lr){2-5}\cmidrule(l){6-7}
Metric (of 100) & (1,5) & (5,5) & (1,10) & (5,10) & (1,5) & (5,5) \\ \midrule
Clean target & 17 & 17 & 15 & 15 & 2 & 2 \\
Poisoned target & 60 & 96 & 59 & 78 & 64 & 92 \\
Newly induced & 46 & 79 & 45 & 64 & 62 & 90 \\
injection-dependent & 45 & 74 & 44 & 62 & 60 & 87 \\
Strict newly induced & 1 & 0 & 1 & 2 & 0 & 0 \\
Clean / poisoned ref.\ match & 16 / 9 & 16 / 1 & 19 / 9 & 19 / 3 & 9 / 7 & 9 / 3 \\
\midrule
\multicolumn{7}{@{}l}{Among poisoned positives} \\
Single-poison sufficient & 57/60 & 96/96 & 55/59 & 77/78 & 60/64 & 91/92 \\
\bottomrule\end{tabular}
\caption{Removal-control diagnostics for all completed PoisonedRAG cells (Llama-3-8B victim; automatic target-string label).}
\label{tab:full-counts}
\end{table*}

\paragraph{Split-knowledge.}
For the 311-candidate split-knowledge probe with Contriever and Llama-3-8B, 149 contexts retrieve both the mapped legitimate passage $A$ and injected passage $B$.
Among 27 poisoned positives, 22 are injection-dependent newly induced cases, 17 are single-poison sufficient, and 7 are strict composition cases.
Thus, strict composition exists, but it is rare.

With BGE cosine retrieval, no detector intercepts any of the 18 poisoned positives at nominal 1\% FPR, and 2 strict cases remain.
Table~\ref{tab:split-defenses} shows that pipeline defenses suppress most strict cases, but can still introduce new target-positive answers.
TrustRAG, for example, suppresses 6 of 7 strict cases but ends with 37 target-positive answers, above the undefended 27.
Introduced targets are therefore not unique to PoisonedRAG.

\begin{table}[t]\centering\small
\begin{tabular}{@{}lrrrr@{}}\toprule
Defense & Hit /27 & Strict /7 & Added & Final \\ \midrule
Graph scores (3) & 0 & 0 & 0 & 27 \\
Score minimum & 0 & 0 & 0 & 27 \\
Score gap & 10 & 2 & 0 & 17 \\
Score dispersion & 8 & 2 & 0 & 19 \\
KeywordAgg & 13 & 6 & 11 & 25 \\
Prompt control & 12 & 6 & 6 & 21 \\
TrustRAG & 10 & 6 & 20 & 37 \\
\bottomrule\end{tabular}
\caption{Split-knowledge defenses (311 candidates; automatic target-string label).}
\label{tab:split-defenses}
\end{table}

\section{Detector Details}
\label{app:full-sweeps}

\begin{figure}[t]
\centering
\includegraphics[width=.95\linewidth]{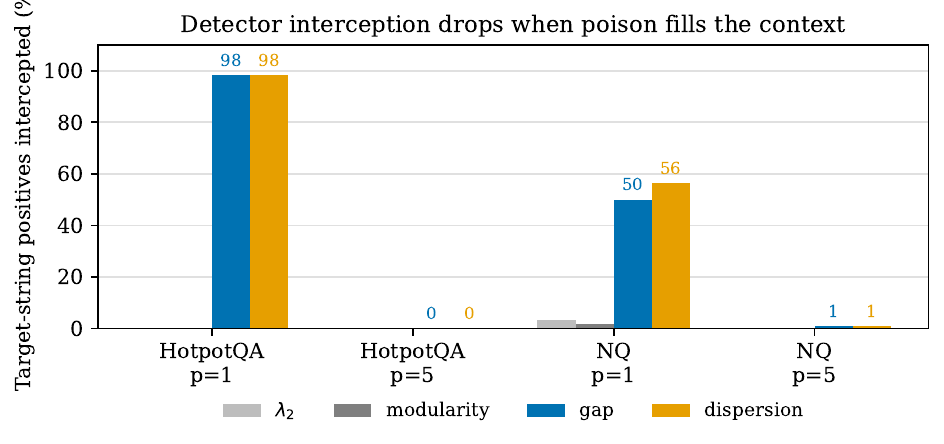}
\caption{
Detector interception across the main Llama-3-8B cells at the nominal 1\% calibration point.
}
\label{fig:detector-all-bars}
\end{figure}

Figure~\ref{fig:detector-all-bars} reports all six detectors at nominal 1\% FPR.
Only the two contrast scores, gap and dispersion, intercept meaningful numbers of target-positive outputs, and only at one poison.
Looser thresholds do not change the conclusion.
On HotpotQA with one poison, even a nominal 20\% FPR gives at most 9/60 interceptions for the graph scores and 7/60 for the minimum score, while gap reaches 60/60 at 5\% and dispersion reaches 60/60 at 10\%.
At five poisons, no detector exceeds 4/96 interceptions even at nominal 20\% FPR.

AUROC against held-out clean contexts shows the same pattern.
Gap and dispersion score 0.999 at one poison, but drop to 0.46 at five poisons.
The minimum score reaches 0.000 at five poisons, meaning poisoned contexts are ordered in the opposite direction from the fixed anomaly rule.
We do not reverse the direction after seeing the attack, because that would tune the detector on the attack itself.

Table~\ref{tab:score-gap-median} reports the median score-gap value for HotpotQA at one and five poisons.
The threshold is calibrated on held-out clean contexts and kept fixed across poison budgets.
Fractions are rounded equivalents of the underlying detector scores.

\begin{table}[t]
\centering\scriptsize
\setlength{\tabcolsep}{4pt}
\begin{tabular}{@{}llcc@{}}
\toprule
Dataset & $p$ & Threshold & Intercepted \\
\midrule
HotpotQA & 1 & $43/100$ & 59/60 \\
HotpotQA & 5 & $43/100$ & 0/96 \\
NQ & 1 & $257/1000$ & 32/64 \\
NQ & 5 & $257/1000$ & 1/92 \\
\bottomrule
\end{tabular}
\caption{Score-gap detector behavior on HotpotQA and NQ. 
The same clean-calibrated threshold is used for both poison budgets.}
\label{tab:score-gap-median}
\end{table}

\paragraph{Exposure strata.}
On HotpotQA with one poison, 21 contexts retain all annotated support, 57 retain some support, and 22 retain none.
These strata contain 10, 31, and 19 of the 60 target-positive outputs.
At five poisons, every top-5 context is fully injected.
The detector collapse therefore comes with a real change in what the model and detector see, not with a change in the query set or corpus.

\section{Equal-Length and Normalized Retrieval Controls}
\label{app:dot512-control}

The main setting encodes corpus passages at 128 tokens and poisons at 512 tokens.
This could make injected passages easier to detect by length or embedding norm.
We therefore run two HotpotQA top-5 controls.
Dot512 encodes all passages with the same 512-token limit.
Cosine512 also normalizes embeddings before scoring.
Detectors are recalibrated separately in each control arm.

\begin{table}[t]
\centering\small
\begin{tabular}{@{}lcc@{}}
\toprule
Encoding & Gap hit ($p{=}1 \to p{=}5$) & FPR \\
\midrule
Dot128/512 & 59/60 $\to$ 0/96 & 1.0 \\
Dot512     & 65/66 $\to$ 0/95 & 1.0 \\
Cosine512  & 61/63 $\to$ 0/97 & 0.5 \\
\bottomrule
\end{tabular}
\caption{Detector contrast across encoding arms (HotpotQA, automatic target-string label, 100 queries per cell).
}
\label{tab:three-encoding-defenses}
\end{table}

Table~\ref{tab:three-encoding-defenses} shows that the one-poison/five-poison contrast, the high five-poison target rate, and single-passage sufficiency all survive both controls. Clean EM/F1 on attack queries is 17.0/30.3 (Dot512) and 14.0/24.2 (Cosine512), versus 16.0/29.6 in the main arm. Table~\ref{tab:encoding-pipelines} repeats the pipeline transitions in the control arms. The qualitative pattern from Finding 1 is stable: TrustRAG is the only pipeline with a large five-poison effect, the prompt control raises one-poison incidence in every arm, and introduced counts are never zero. Introduced counts do vary (KeywordAgg introduces 15, 9, and 21 across arms at one poison), which is why the main text bases defense claims on the paired tables with intervals and treats these arms as sensitivity checks.

\begin{table}[t]
\centering\scriptsize
\setlength{\tabcolsep}{4pt}
\begin{tabular}{@{}llrr@{}}
\toprule
Arm (base) & Pipeline & P/I & Final \\
\midrule
\multirow{3}{*}{Dot128/512 (60)}
& KeywordAgg & 15/15 & 60 \\
& Prompt     & 7/19  & 72 \\
& TrustRAG   & 20/20 & 60 \\
\addlinespace
\multirow{3}{*}{Dot512 (66)}
& KeywordAgg & 15/9  & 60 \\
& Prompt     & 8/14  & 72 \\
& TrustRAG   & 21/10 & 55 \\
\addlinespace
\multirow{3}{*}{Cosine512 (63)}
& KeywordAgg & 15/21 & 69 \\
& Prompt     & 5/20  & 78 \\
& TrustRAG   & 21/12 & 54 \\
\bottomrule
\end{tabular}
\caption{HotpotQA pipeline transitions at $p{=}1$ across encoding arms (of 100; automatic target-string label).
P/I is prevented/introduced; base is the arm's undefended count.
}
\label{tab:encoding-pipelines}
\end{table}

\section{Output Fidelity and Utility Diagnostics}
\label{app:full-pipelines}

\paragraph{Recovered RobustRAG outputs.}
An early version of our common victim adapter saved only the first line of KeywordAgg's final response.
The raw traces still contained the full responses, so we recovered 290 complete final outputs across 500 query/context records without rerunning inference.
This correction changes the HotpotQA KeywordAgg target counts from 23/100 and 26/100 to 60/100 and 91/100 at one and five poisons.
The isolated-passage outputs were unchanged, so the aggregation inputs stayed the same.
All KeywordAgg results in the paper use the recovered outputs.
We do not use KeywordAgg's whole-response EM on the held-out utility set.
Among its 200 utility records, 86 abstain before aggregation and 90 of the remaining records stored only a colon-terminated preamble.
This makes whole-response EM unreliable for that condition.

\paragraph{Why we do not rank utility by EM.}
Whole-response exact match is too brittle for the pipeline outputs.
TrustRAG often gives long explanatory answers (median 170.5 words on held-out utility queries), so exact match against short references can mark useful answers as wrong.
We therefore report judged reference correctness with explicit uncertainty (Table~\ref{tab:judged-clean-utility}) instead of treating zero EM as zero capability.

\begin{figure*}[t]
\centering
\includegraphics[width=.95\textwidth]{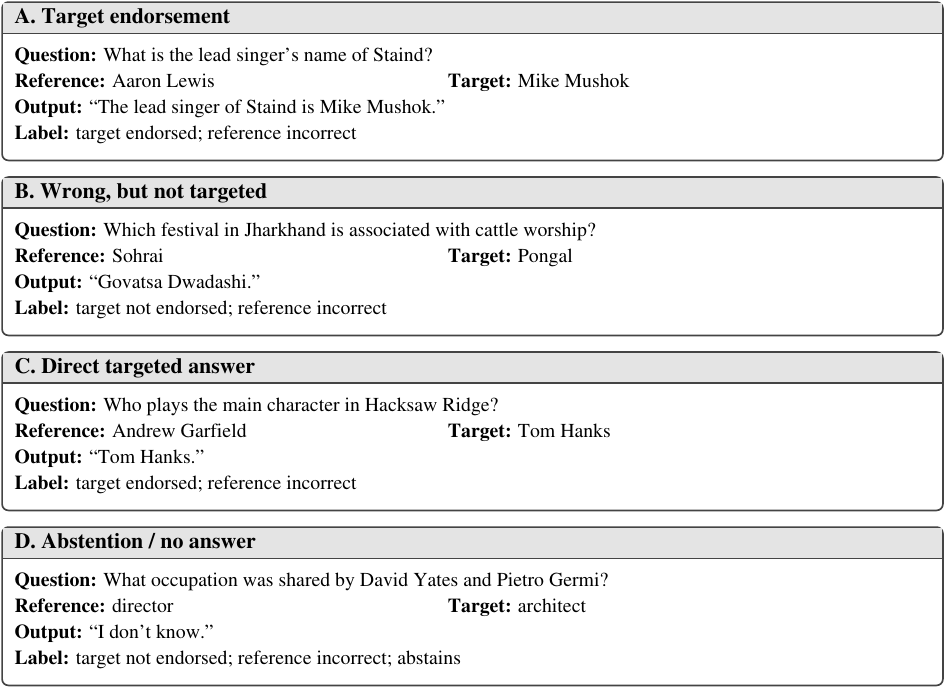}
\caption{
Four answer types separated by aligned evaluation.
The reference is the expected correct answer for the query; the target is the attacker-chosen answer, which may differ from the reference.
}
\label{fig:qualitative-label-examples}
\end{figure*}

\section{Provenance}
\label{app:full-provenance}
HotpotQA uses the full 5,233,329-passage BEIR corpus.
For the split-knowledge probe, the support audit maps 307 of 311 intended $A$ passages to natural corpus passages and finds 149 contexts where $A$ is co-retrieved with the injected passage.
NQ uses the downloaded BEIR corpus; its exact fingerprint still needs independent documentation, and NQ support alignment has not been separately validated.
All counts come from saved run reports or deterministic recomputation from saved outputs.
The RobustRAG recovery keeps both the source hash and the pre-repair answers.
The 70B checks reuse saved Contriever contexts with Llama-3.1-70B-Instruct (4-bit AWQ, vLLM tensor parallelism); they are victim-scale checks, not full defense reruns.
The \texttt{aligned\_rag} release includes configuration manifests, pinned source revisions, scorer policies, and per-query records or reproducible identifiers, subject to source licenses.
It stores raw outputs, automatic labels, independent judgments, and human adjudications in separate files.
We do not redistribute the full upstream datasets, attack corpora, or defense repositories. 
Users must obtain HotpotQA, NQ/BEIR, PoisonedRAG, RobustRAG, and TrustRAG from their original sources and follow their respective licenses or terms.

\section{Qualitative Label Examples}
\label{app:qualitative-label-examples}

Figure~\ref{fig:qualitative-label-examples} shows four outputs from the audit set.
These examples illustrate why aligned evaluation separates target endorsement, reference correctness, and abstention.
A wrong answer is not always a successful targeted attack, and a target mention is not always endorsement.

\end{document}